\documentclass[12pt]{article}
\usepackage{url}
\usepackage{graphicx}
\usepackage{amsfonts,amssymb,amsmath}
\usepackage{float}

\graphicspath{{../figures/}}

\begin{document}

\onecolumn

%Title of paper
\section*{Zeeman-Hyperfine Measurements of a Pseudo-Degenerate Quadruplet in CaF$_2$:Ho$^{3+}$} 

\begin{center}
{Kieran M. Smith$^{1,2,3}$} 
{Michael F. Reid$^{1,2,*}$} 
{Jon-Paul R. Wells$^{1,2, \dag}$}

\bigskip

$^1${School of Physical and Chemical Sciences, University of Canterbury, PB 4800, Christchurch 8140, New Zealand}\\
$^2${The Dodd-Walls Centre for Photonic and Quantum Technologies, New Zealand}\\
$^3${Research School of Physics and Engineering, The Australian National University, Canberra 0200, Australia}\\

\smallskip

$^*$Email: mike.reid@canterbury.ac.nz \\
$\dag$Email: jon-paul.wells@canterbury.ac.nz
 
\bigskip

%Version06: 
%Submitted: 
\today 
\end{center}

\bigskip\noindent \textbf{Abstract---}  We report Zeeman infra-red spectroscopy of electronic-nuclear levels of $^5$I$_8 \rightarrow ^5$I$_7$ transitions of  Ho$^{3+}$ in the C$_{\rm 4v}$(F$^-$) centre in CaF$_2$  with the magnetic field along the $\langle 111\rangle$ direction of the crystal.  Transitions to the lowest  $^5$I$_7$ state, an isolated electronic doublet, and the next group of states, a pseudo-quadruplet consisting of a doublet and two nearby singlets, exhibit strongly non-linear Zeeman splittings and intensity variations. Simulated spectra based upon a crystal-field analysis give an excellent approximation to the data, illustrating the strong predictive ability of the parametrised crystal-field approach. Anti-crossings in the hyperfine splittings, the basis of quantum information storage in rare-earth doped insulating dielectrics, are also predicted.

\bigskip\noindent 
Keywords: holmium; rare earth; Zeeman; spectroscopy; crystal-field

%\bigskip\noindent
%PACS: {71.70.Ch,76.30.Kg,31.15.A-,31.15.bu} 

%\bigskip\noindent
%Version02: 
%\today 

\newpage
%\twocolumn

\section{Introduction}

It is a pleasure to celebrate Professor Marina Popova's 80$^{\rm th}$ birthday by presenting an investigation into highly non-linear Zeeman interactions within the complex electron-nuclear sub-level structure of the Ho$^{3+}$ ion doped into CaF$_{2}$ crystals. We feel that this work is appropriate both because Professor Popova's research group has been amongst the world leaders in hyperfine spectroscopy of rare-earth ions and because it is based upon a most fruitful earlier collaboration \cite{Wells2004}. 

Crystals doped with rare-earth ions are promising candidates for quantum-information devices. Coherent storage of several hours has been demonstrated using the magnetic-hyperfine structure of Eu$^{3+}$ ions in Y$_2$SiO$_5$ \cite{zhong_optically_2015,Ma_YSO_Eu_comb_2021} whilst a coherence time exceeding one second has been demonstrated in Er$^{3+}$:Y$_2$SiO$_5$
using a 7 Tesla magnetic field \cite{rancic2018}. The long coherence times recorded in Ref. \cite{zhong_optically_2015} made use of ZEFOZ, or ``clock'' transitions, which are insensitive to small magnetic fluctuations. These transitions are generally associated with anticrossings (avoided crossings) of hyperfine sublevels. Importantly, we have recently demonstrated that it is possible to perform crystal-field calculations for the C$_{\rm 1}$ point group symmetry sites in Y$_{2}$SiO$_{5}$ \cite{Horvath2019} and exploit their predictive ability to interpret hyperfine spectra \cite{Mothkuri_zeeman_2021}. 

The investigation of magnetic-hyperfine structure in crystals doped with ions such as Eu$^{3+}$ and Er$^{3+}$ requires high-resolution laser spectroscopy \cite{macfarlane1987}. On the other hand,  Ho$^{3+}$ has a large nuclear magnetic moment, and hyperfine splittings may be investigated using conventional techniques. The earliest optical studies of Ho$^{3+}$ hyperfine structure date back to the pioneering investigations of Gerhard Dieke at Johns Hopkins University in the mid 1960s. This work, using a 5 metre vacuum spectrometer custom built by Jarrell-Ash, presented fully resolved hyperfine structure and Zeeman splittings for the Z$_{1}\longrightarrow$J$_{1}$ transition at 23918 cm$^{-1}$ in LaCl$_{3}$ crystals \cite{dieke1964,crosswhite1967}. In this work we present Zeeman spectroscopy of Ho$^{3+}$ in the C$_{\rm 4v}$(F$^{-}$) centre in CaF$_2$ crystals, also using conventional absorption spectroscopy. We have previously reported the observation of an exceedingly complex hyperfine structure arising from the magnetic interactions of a two singlets in close proximity to an orbital doublet in the excited state convolved with ground state hyperfine structure arising from two close lying singlets coupled by a pseudo-quadrupole interaction \cite{Wells2004}. There, we obtained excellent agreement between theory and experiment for both the energy levels and the transition intensities for the spectrum measured without the application of a magnetic field. Here we demonstrate the ability of the crystal-field model to {\it predict} the Zeeman splittings of the hyperfine states and their transition intensities, including the presence of anti-crossings - the effect upon which the ZEFOZ technique is based. \\

\section{Experimental}

Calcium fluoride crystals containing 0.01 molar percent of HoF$_{3}$ were grown under vacuum, using the Bridgman-Stockbarger technique in a 38 kW RF furnace. The crystal charge was placed in a graphite crucible (together with a small quantity of PbF$_{2}$ to act as an oxygen scavenger) and lowered through the temperature gradient provided by the furnace work coil over five days. The crystal was cleaved along the $\langle 111 \rangle$ plane to allow the desired orientation in the magnet. The absorption length was approximately 5\,mm. 

Infrared spectroscopy was performed using a 0.075\,cm$^{-1}$ resolution Bruker Vertex 80 with an optical path purged by N$_2$ gas.  Zeeman spectroscopy
was performed using a 4\,T, simple solenoid, superconducting magnet with samples mounted into a copper sample holder fixed through the centre of the solenoid. Measurements were carried out at 4.2\,K since the copper sample mount is in direct thermal contact with the liquid helium bath. 

\section{Theoretical Background}

The Hamiltonian appropriate for modelling the $4f^N$ configuration is
\cite{Carnall1989,GoBi96,liu_electronic_2006,Reid2016}
\begin{equation}
  H = H_{\mathrm{FI}} + H_{\mathrm{CF}}  + H_{\mathrm{HF}} + H_{\mathrm{Z}} .
  \label{eqn:hdefn}
\end{equation} 
The terms in this equation represent
the free-ion contribution, the crystal-field interaction, the  electron-nuclear hyperfine interaction, and the Zeeman interaction. \\

% The free-ion Hamiltonian  may be written as
% \begin{align}
%  H_{\mathrm{FI}} &= E_\text{avg} + \sum_{k=2,4,6} F^k f_k + \zeta _{A_{\mathrm{SO}}}
%  \nonumber\\
%  &+\alpha L(L+1) + \beta G(G_2) + \gamma G(R_7) 
%                  + \sum_{i = 2,3,4,6,7,8} T^i t_i
%  \nonumber \\  
%  &  + \sum_{i=0,2,4} M^i m_i + \sum_{i=2,4,6} P^i p_i . 
%   \label{eqn:h_fi_defn}
% \end{align}
% $E_\text{avg}$ is a constant configurational shift, $F^k$ Slater
% parameters characterizing aspherical electrostatic repulsion, and
% $\zeta$ the spin-orbit coupling constant. The other terms 
% parametrize two- and three- body interactions, as well as higher-order
% spin-dependent effects
% \cite{Carnall1989,liu_electronic_2006}.

The free-ion part of the Hamiltonian has been discussed in great detail in the above references. 
The crystal-field Hamiltonian has the form 
\begin{equation}
    H_{\mathrm{CF}} = \sum_{k,q} B^{k}_q C^{(k)}_q , 
\end{equation}
with k = 2, 4, 6, q = -k,..,k. The $B^{k}_q$ are crystal field parameters, $C^{(k)}_q$ are spherical tensor operators. In C$_{\rm 4v}$  symmetry we may choose the $z$ axis of the site to be the four-fold axis, so only parameters with $q=0$, $\pm 4$ are non-zero.  

The hyperfine Hamiltonian $H_{\mathrm{HF}}$ is discussed in detail in Ref.~\cite{Wells2004}.  The holmium nucleus has a spin $I=7/2$, and the electronic states are coupled to the nuclear spins by the hyperfine interaction, giving 8 electronic-nuclear states for each electronic level. 
%We use  basis states $|JM_J,IM_I\rangle$ in our calculations.  
For holmium the magnetic-hyperfine interaction is much larger than the nuclear-quadrupole interaction, but both are included in the calculation. It is important to note that for singlet electronic states the diagonal matrix elements of the hyperfine operators are zero, so the splittings are a result of coupling to nearby electronic states. 

The effect of an external magnetic field is given by the Zeeman Hamiltonian
\begin{equation}
    %H_{\mathrm{Z}} = \mu_B\sum_{i}^{N} B (l_i +2s_i)
    H_{\mathrm{Z}} = \mu_B\mathbf{B} \cdot (\mathbf{L} +\mathbf{2S}) , 
  \label{eqn:Zeeman}
\end{equation}
where B is the  magnetic field. 

CaF$_2$ is a cubic crystal. In the C$_{\rm 4v}$ symmetry centres studied here a Ca$^{2+}$ ion is replaced by a Ho$^{3+}$ ion with a nearest neighbour interstitial F$^-$ ion providing charge compensation (this has generally been denoted as the C$_{\rm 4v}$(F$^{-}$) centre in the literature) \cite{Wells2004}.  The four-fold axis of the site can be along any of the three crystal axes. In this study we orient the crystal so that the magnetic field can be applied along the $\langle 111 \rangle$ axis. Thus the magnetic field is along the body diagonal direction and equally inclined at an angle of ${\rm cos}^{-1}(1/\sqrt{3})=54.7^{~\circ}$ to the three C$_{\rm 4v}$ centre orientations. As a result, each site experiences the same magnetic splitting, greatly simplifying the spectra. The magnetic field also reduces the symmetry experienced by the Ho$^{3+}$ ions to C$_{1}$. 

In our previous work \cite{Wells2004} we performed a detailed study that made use of laser spectroscopy and IR spectroscopy. Our crystal-field modelling gave excellent agreement with the electronic and hyperfine energies. Magnetic and electric dipole transition intensities \cite{reid_transition_2006} were also calculated, again with excellent agreement. We use the same crystal-field and transition-intensity parameters as in that work. The only addition to the Hamiltonian is the Zeeman interaction, Eq.~(\ref{eqn:Zeeman}).  Theoretical spectra are generated using a Lorentzian line shape with a full-width half-maximum of 0.09 cm$^{-1}$. The relative population of the  Z$_1\gamma_1$ and  $Z_2\gamma_2$ states of the $^5$I$_8$ multiplet was calculated according to the Boltzmann distribution.

\section{Results and Discussion}

\begin{figure}[tb!] % Zero Field Spectra
\centering
\includegraphics[width=0.6\columnwidth]{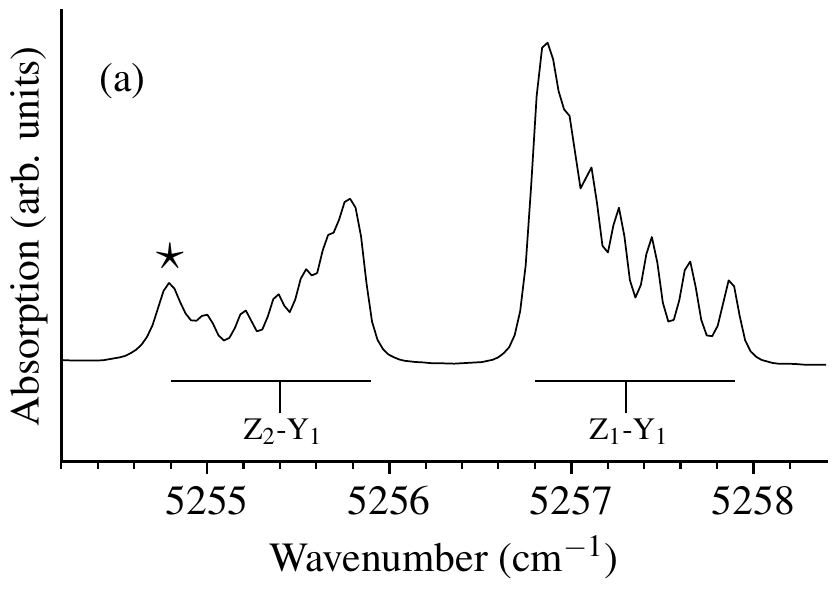}\\
\includegraphics[width=0.6\columnwidth]{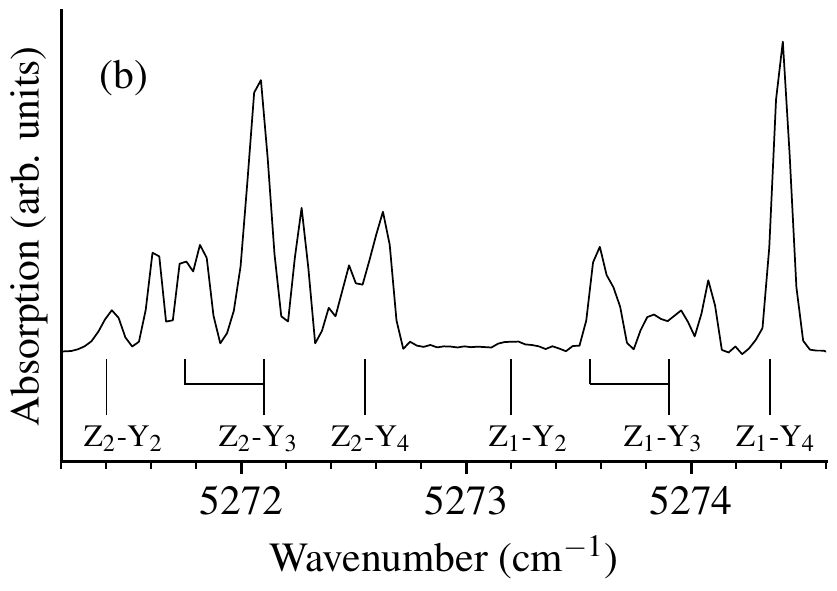}
\caption{\label{fig:zerofield}  
4.2~K spectra at zero magnetic field for (a) transitions from Z$_1\gamma_1$, $Z_2\gamma_2$  of $^5$I$_{8}$ to  Y$_1\gamma_5$ of  $^5$I$_{7}$,  
(b) transitions from Z$_1\gamma_1$, $Z_2\gamma_2$  of $^5$I$_{8}$ to the $Y_2\gamma_3$,  $Y_3\gamma_5$, $Y_4\gamma_2$ states of  $^5$I$_{7}$.
The feature at 5254.8\,cm$^{-1}$, labelled by an asterisk, is an atmospheric H$_{2}$O absorption line. 
}
\end{figure}

The ground-state levels of the C$_{\rm 4v}$(F$^{-}$) centre in CaF$_{2}$:Ho$^{3+}$ are comprised of two singlets separated by 1.7 cm$^{-1}$ \cite{mujaji1992,martin1993}.
As singlets (transforming as the $\gamma_1$ and $\gamma_2$ irreps of the C$_{\rm 4v}$ single group) neither of these states posses a magnetic moment. However, they can acquire one via the pseudo-quadrupole interaction (and concomitant $A_{8}I_{z}J_{z}$ mixing) between them. Figure \ref{fig:zerofield}(a) shows a 4.2~K, zero magnetic field spectrum of the transitions from these ground state levels to the lowest level of the $^{5}$I$_{7}$ multiplet at approximately 5257 cm$^{-1}$ which is an orbital doublet (Y$_{1}\gamma_{5}$). The intensity distribution amongst the levels is governed by the wavefunction admixtures and the Boltzmann population distribution amongst the hyperfine levels \cite{Wells2004}. Figure \ref{fig:zerofield}(b) shows the 4.2~K, zero magnetic field spectrum of transitions from the ground-state pseudo doublet to another excited state doublet level (Y$_{3}\gamma_{5}$) at 5273.9 cm$^{-1}$. In this case the spectrum is exceedingly complex. This arises due to the close proximity of two singlets (Y$_{2}\gamma_{3}$ at 5273.1 cm$^{-1}$ and Y$_{4}\gamma_{2}$ at 5274.4 cm$^{-1}$). Thus the pattern observed arises from the mixing of the singlet states with the doublet via the perpendicular hyperfine ($A_{\perp}\frac{1}{2}(I_{+}J_{-}+I_{-}J_{+})$) interaction. Due to the strong wavefunction mixing, there is considerable redistribution of intensities and $I_{z}$ ceases to be a good quantum number, with the $\Delta I_{z} = 0$ selection rule breaking down. Electronic selection rules also break down. Notably, transitions from Z$_1\gamma_1$ to $Y_2\gamma_3$ becomes allowed due to mixing via the hyperfine interaction. 

\begin{figure}[tb!] % Energies
\centering
\includegraphics[width=0.5\columnwidth]{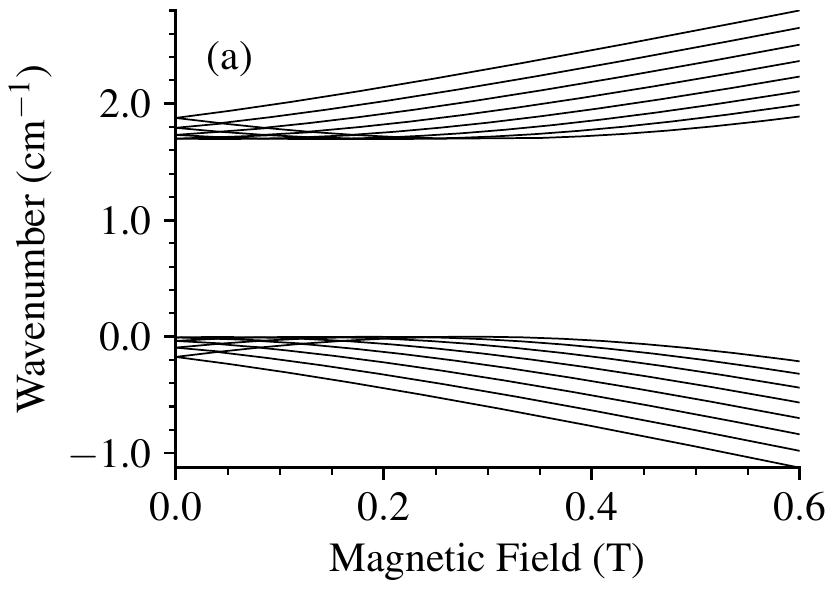}\\
\includegraphics[width=0.5\columnwidth]{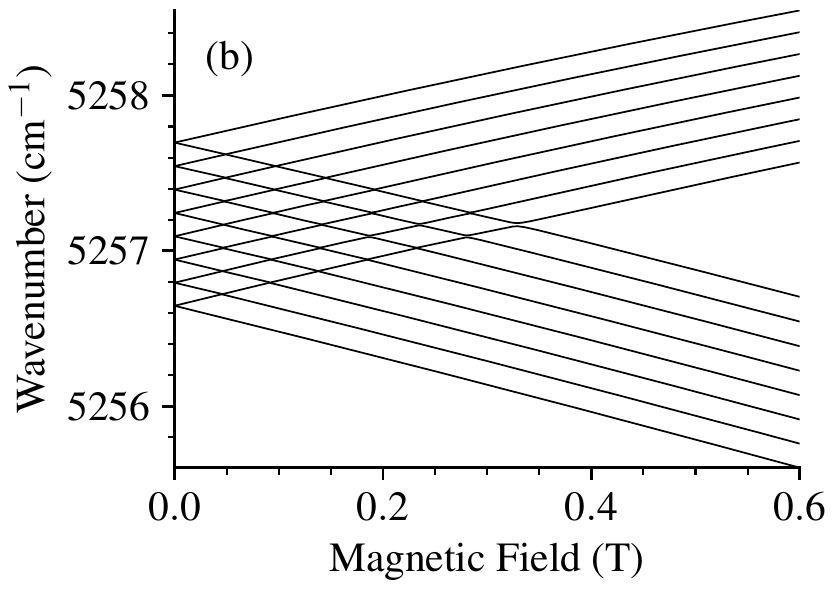}\\
\includegraphics[width=0.5\columnwidth]{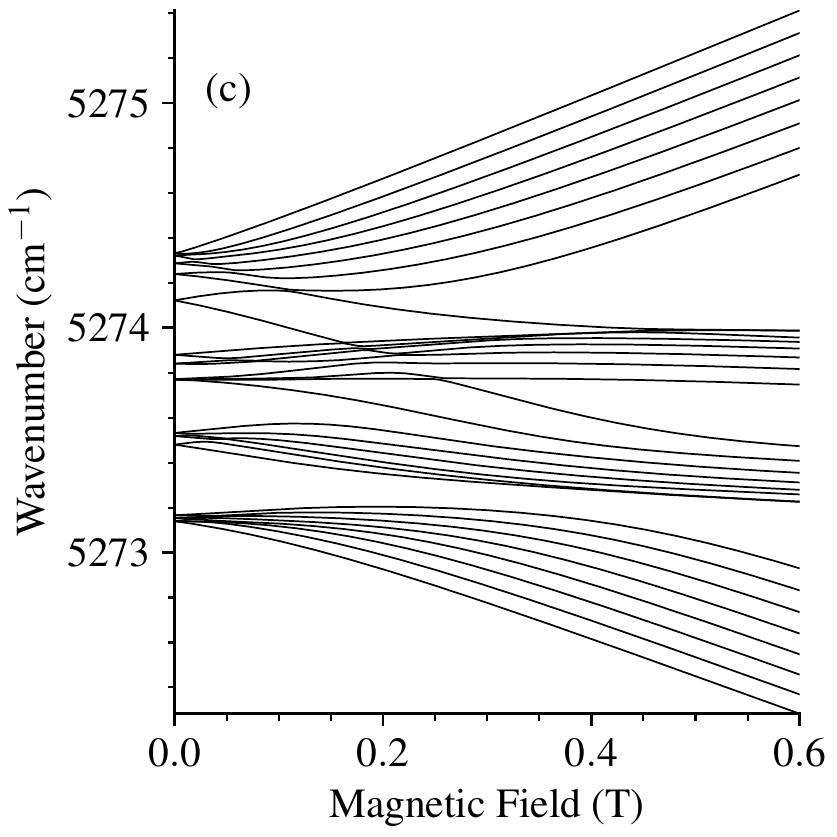}
\caption{\label{fig:Energies}  
Calculated Zeeman-hyperfine energies for a magnetic field along the $\langle 111 \rangle$ direction for (a) Z$_1\gamma_1$, $Z_2\gamma_2$  of $^5$I$_{8}$,  
(b) Y$_1\gamma_5$ of  $^5$I$_{7}$, (c) $Y_2\gamma_3$,  $Y_3\gamma_5$, $Y_4\gamma_2$ of  $^5$I$_{7}$. }
\end{figure}

Figure \ref{fig:Energies}(a) plots the calculated $\langle 111 \rangle$ Zeeman splittings for the ground state pseudo doublet. This splitting is governed by the quadratic Zeeman effect between the singlets as they repel each other in the applied field. Our calculations agree with EPR measurements, giving $g_\parallel =14.8$ ($g_\perp$ is zero) \cite{kornienko1972}.
Figures \ref{fig:Energies}(b) and (c) plot the calculated $\langle 111 \rangle$ Zeeman splittings for the isolated Y$_{1}$ doublet and the more complex pseudo-quadruplet pattern comprising the Y$_{2,3,4}$ states. The former exhibits a linear Zeeman effect, as would be expected, whilst the later shows strongly non-linear splittings. In the absence of the hyperfine interaction the two singlets would not interact with each other and the pattern would be entirely governed by two effects: (1) the first order Zeeman effect within the doublet and (2) the quadratic Zeeman effect between the doublet and the two singlets. However, the perpendicular hyperfine interaction creates strong wavefunction admixtures, which alters the character of the electronic levels, adding to the complexity of the splittings. More generally, it is the extreme sensitivity of these effects that makes this system useful as a test of the predictive ability of the crystal-field model.

% Y1 Transitions
\begin{figure}[tb!]
\centering
\includegraphics[width=0.45\columnwidth]{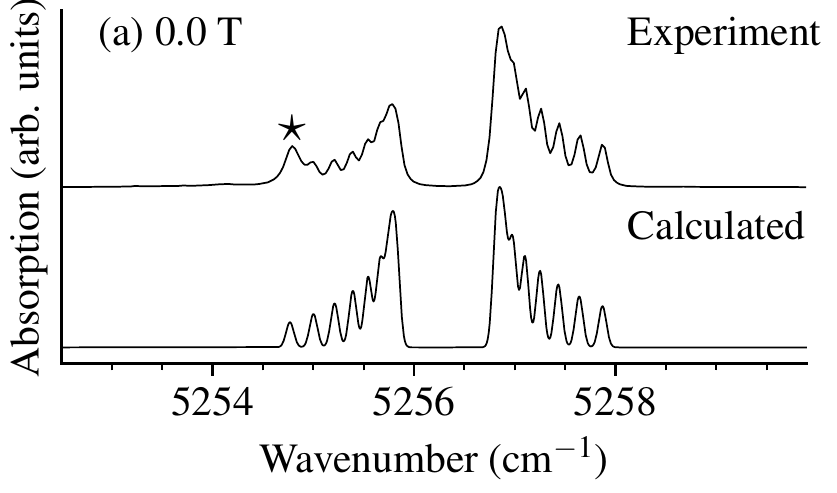}\\
\includegraphics[width=0.45\columnwidth]{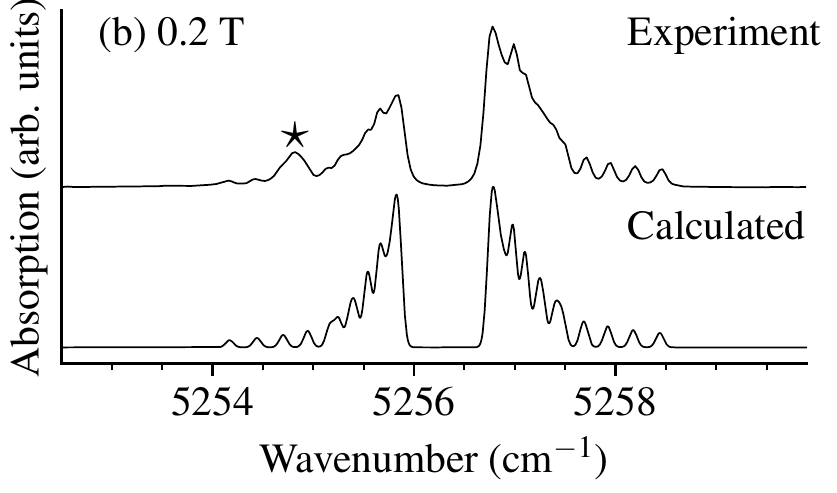}\\
\includegraphics[width=0.45\columnwidth]{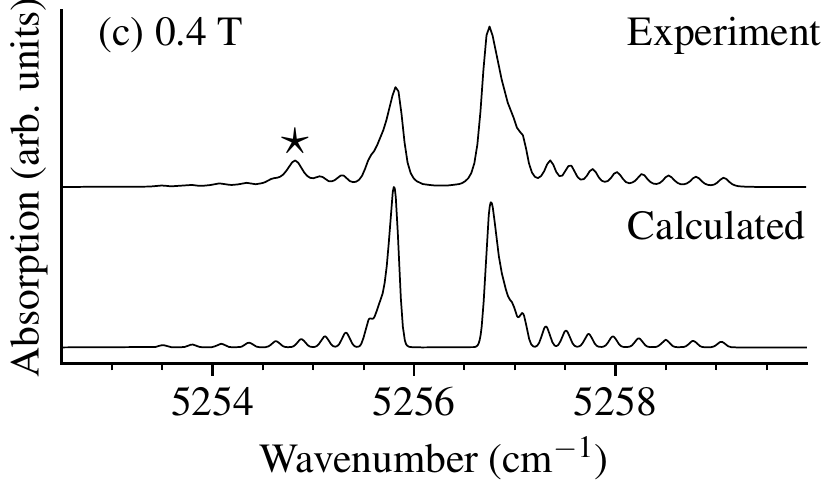}\\
\includegraphics[width=0.45\columnwidth]{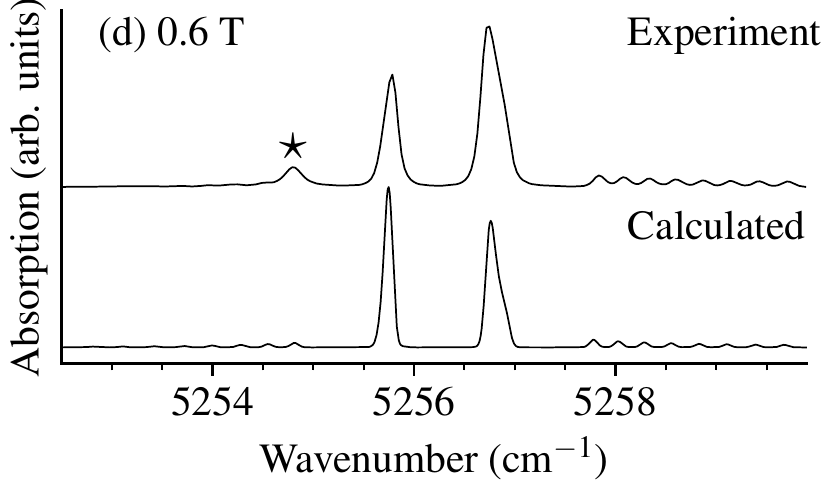}\\
\caption{\label{fig:Z1Z2Y1}  
Experimental and calculated 4.2~K $\langle 111 \rangle$ Zeeman spectra for the Z$_{1,2}\gamma_{1,2}\longrightarrow$Y$_1\gamma_5$ transitions with (a) 0.0\,T, (b) 0.2\,T, (c) 0.4\,T, (d) 0.6\,T.
}
\end{figure}

\begin{figure}[tb!] % Y1 Transitions - maps
\centering
\includegraphics[width=0.8\columnwidth]{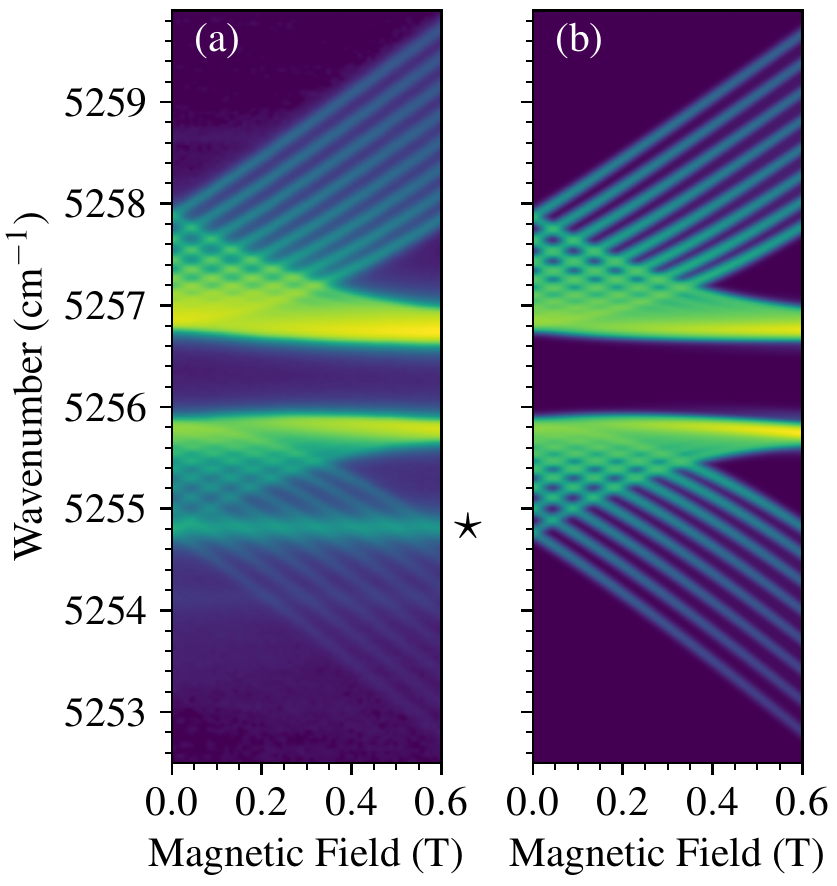}
\caption{\label{fig:Z1Z2Y1map}  
Experimental (a) and calculated (b) 4.2~K $\langle 111 \rangle$ Zeeman spectra for the Z$_{1,2}\gamma_{1,2}\longrightarrow$Y$_1\gamma_5$ transitions for magnetic fields up to 0.6~T.}
\end{figure}

Experimental and calculated Zeeman spectra for the Z$_{1,2}\gamma_{1,2}\longrightarrow$Y$_1\gamma_5$ transitions are given in Figure \ref{fig:Z1Z2Y1} for 0.0\,T 0.2\,T 0.4\,T, and 0.6\,T, and experimental and calculated maps from 0.0\,T to  0.6\,T are given in Figure \ref{fig:Z1Z2Y1map}. These transitions are similar to the singlet to doublet transition shown in Figures 1 and 3 of Ref.~\cite{Boldyrev2019}.  However, the larger hyperfine and magnetic splittings of our singlet state makes our spectra more complex. 

The application of a magnetic field along the $\langle 111 \rangle$ direction reduces the symmetry to C$_1$, and in principle all transitions should be allowed. However, transitions violating the  $\Delta I_{z} = 0$ selection rule are not observed, and the allowed transitions are as shown in Table 8 of Ref.~\cite{Wells2004}. As the magnetic field is increased, the Y$_1\gamma_5$ electronic doublet splits in two. Transitions from  Z$_1\gamma_1$ to the higher Y$_1\gamma_5$ states increase in spacing, whereas transitions from Z$_1\gamma_1$ to the lower Y$_1\gamma_5$ states tend towards the same energy, since in the latter case the shift in energy of the states as a function of magnetic field is similar (see Figures \ref{fig:Energies}(a) and \ref{fig:Energies}(b)). 
The calculations give excellent agreement with experiment, not only for the transition energies, but also the transition intensities. 

% Y234 transitions
  
\begin{figure}[tb!]
\centering
\includegraphics[width=0.5\columnwidth]{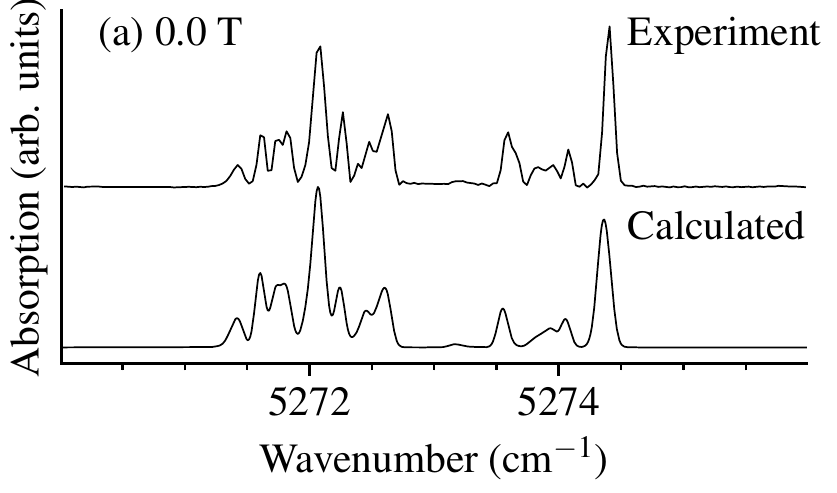}\\
\includegraphics[width=0.5\columnwidth]{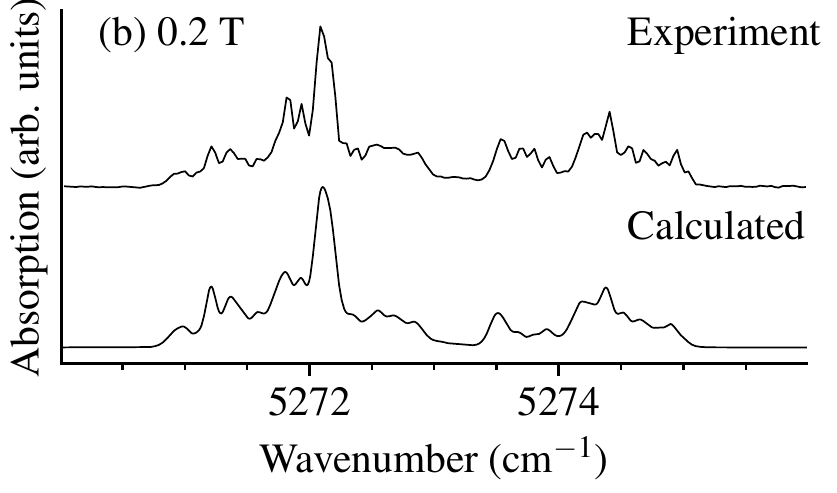}\\
\includegraphics[width=0.5\columnwidth]{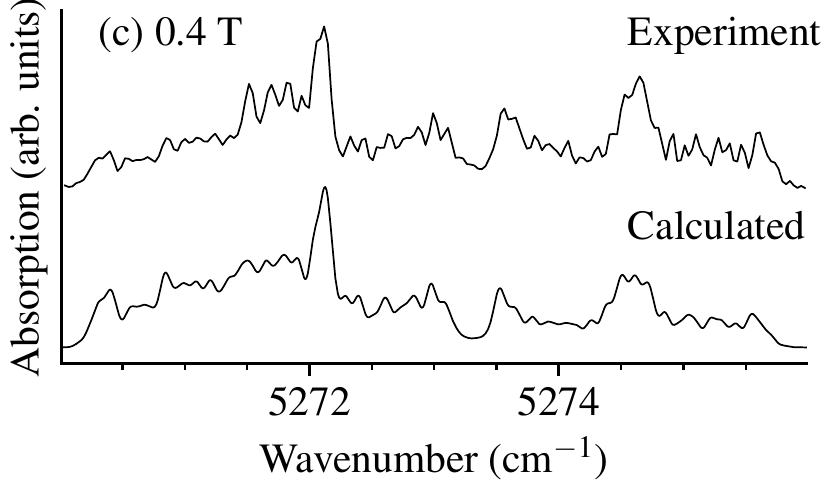}
\caption{\label{fig:Z1Z2Y2Y3Y4}  
Experimental and calculated 4.2~K $\langle 111 \rangle$ Zeeman spectra for the Z$_{1,2}\gamma_{1,2}\longrightarrow$Y$_{2,3,4}\gamma_{3,5,2}$ transitions with (a) 0.0\,T, (b) 0.2\,T, (c) 0.4\,T.
}
\end{figure}

\begin{figure}[tb!] % Y234 transitions - maps
\centering
\includegraphics[width=0.8\columnwidth]{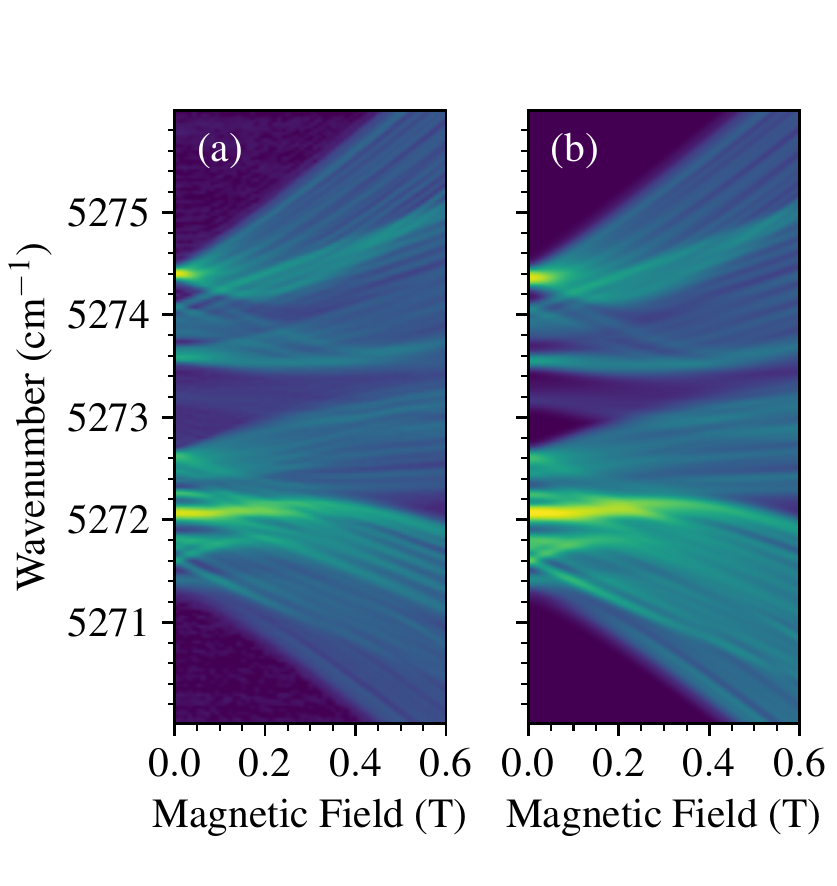}
\caption{\label{fig:Z1Z2Y2Y3Y4map}  
Experimental (a) and calculated (b) 4.2~K $\langle 111 \rangle$ Zeeman spectra for the Z$_{1,2}\gamma_{1,2}\longrightarrow$Y$_{2,3,4}\gamma_{3,5,2}$ transitions for applied magnetic fields up to 0.6~T.
}
\end{figure}

Experimental and calculated Zeeman spectra for the Z$_{1,2}\gamma_{1,2}\longrightarrow$Y$_{2,3,4}\gamma_{3,5,2}$ transitions are given in Figure \ref{fig:Z1Z2Y2Y3Y4} for 0.0\,T,  0.2\,T, and 0.4\,T, and experimental and calculated maps from 0.0\,T to  0.6\,T are given in Figure \ref{fig:Z1Z2Y2Y3Y4map}.
Here we choose comparatively low field data (fields up to four Tesla were measured) to illustrate the effect, since at significantly higher fields 
%(for the ground state $3g^{2}_{(111)}=g^{2}_{\parallel}+2g^{2}_{\perp}$=8.56 %\cite{kornienko1972}) 
the measured spectrum is too widely spread to observe details and the intensity lowers significantly. As can be seen in either figure, there is astonishingly good agreement between the simulated spectra and the experimental data, both for the energies and the transition intensities.

\begin{figure}[tb!]
\centering
\includegraphics[width=0.5\columnwidth]{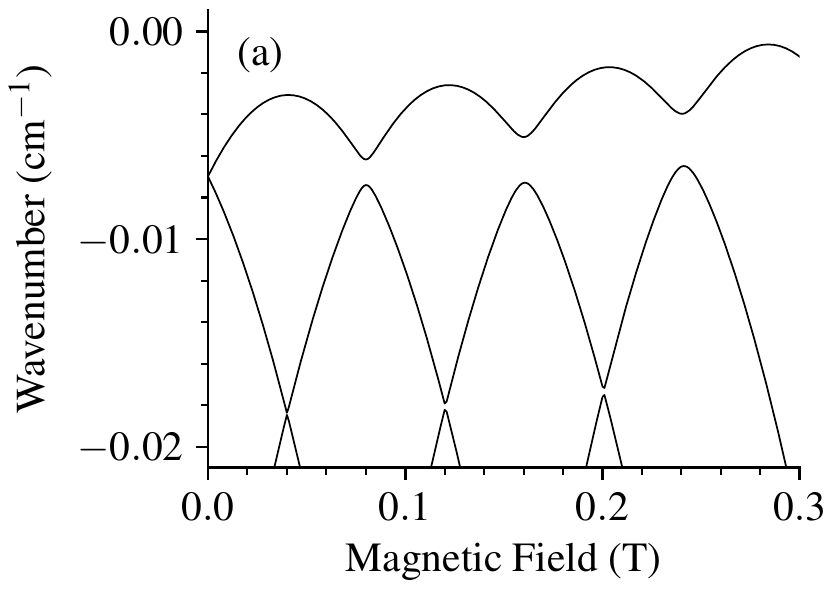}\\
\includegraphics[width=0.5\columnwidth]{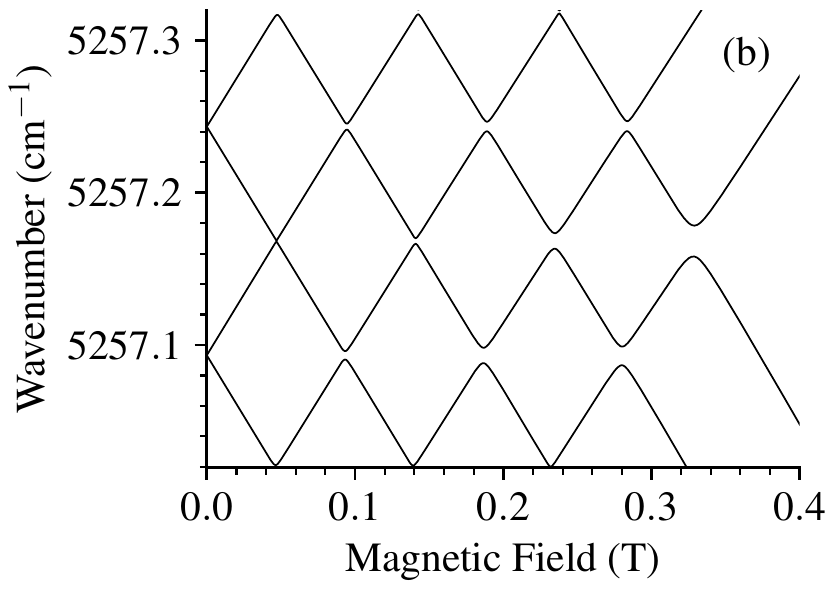}
\caption{\label{fig:Anticrossings} 
Expanded energy-level calculations with a magnetic field along the $\langle 111 \rangle$ direction for the (a) Z$_1\gamma_1$ of $^5$I$_{8}$ and  
(b) Y$_1\gamma_5$ of  $^5$I$_{7}$, electronic states.}
\end{figure}

Figure \ref{fig:Anticrossings} shows calculated anticrossings within the Z$_1\gamma_1$ and Y$_1\gamma_5$ states, enlarged from Figures \ref{fig:Energies}(a) and  \ref{fig:Energies}(b).  The largest splitting for the anticrossings in Y$_1\gamma_5$ are comparable with those in Figure 3 of Ref.~\cite{Boldyrev2019} (0.06\,cm$^{-1}$). However, the larger line-widths in CaF$_2$ compared to LiYF$_4$ would make resolving the anticrossings difficult, even for a low-concentration crystal and higher-resolution measurement, such as those of Figure 5 of Ref.~\cite{Wells2004}.  Anticrossings within Z$_1\gamma_5$ are of the order of 70 to 150\,MHz, and could potentially be studied at RF frequencies. 
Enhancement of the coherence time at anticrossings between hyperfine levels of
the Z$_1$ and Z$_2$ electronic states of holmium in a molecular magnet has been demonstrated in pulsed EPR measurements \cite{Shiddiq2016}. Similar measurements would be possible for the C$_{4v}$(F$^-$) CaF$_2$ centre. However, the
splitting between Z$_1\gamma_1$ and Z$_2\gamma_2$ is much larger (50.8 GHz) than the 9 GHz splitting utilized in Ref.\ \cite{Shiddiq2016}.
There are clearly numerous anticrossings in the Y$_2\gamma_3$, Y$_3\gamma_5$, Y$_4\gamma_2$ states. However, the complexity of the spectra makes definitive identification extremely difficult. 

\section{Conclusions}

We have investigated the hyperfine structure of Ho$^{3+}$  ions in the C$_{\rm 4v}$(F$^-$)  centre in CaF$_2$, under the influence of a magnetic field applied along the $\langle 111 \rangle$ crystallographic direction. In particular, we have presented results for transitions to an isolated orbital doublet and a pseudo-quadruplet grouping consisting of a doublet and two nearby singlets. A crystal-field model developed previously \cite{Wells2004}, with the addition of only the Zeeman term to the interaction Hamiltonian, accounts for the experimental energies and transition intensities with remarkable precision. This bodes well for the more challenging task of predicting ZEFOZ points in the low symmetry materials used for quantum information storage.

\clearpage 
%\bibliographystyle{physss}
%\bibliographystyle{elsarticle-num}
%\bibliography{hocaf2,hoyso,ysozeeman,eryso_crystal_field}

\begin{thebibliography}{10}

\bibitem{Wells2004}
J.-P.~R. Wells, G.~D. Jones, M.~F. Reid, M.~N. Popova, and E.~P. Chukalina.
\newblock Mol. Phys. \textbf{102}, 1367 (2004).

\bibitem{zhong_optically_2015}
M.~Zhong, M.~P. Hedges, R.~L. Ahlefeldt, J.~G. Bartholomew, S.~E. Beavan, S.~M.
  Wittig, J.~J. Longdell, and M.~J. Sellars.
\newblock Nature \textbf{517}, 177 (2015).

\bibitem{Ma_YSO_Eu_comb_2021}
Y.~Ma, Y.-Z. Ma, Z.-Q. Zhou, C.-F. Li, and G.-C. Guo.
\newblock Nature Communications \textbf{12}, 2381 (2021).

\bibitem{rancic2018}
M.~Ran{\v c}i{\'c}, M.~P. Hedges, R.~L. Ahlefeldt, and M.~J. Sellars.
\newblock Nature Physics \textbf{14}, 50 (2018).

\bibitem{Horvath2019}
S.~P. Horvath, J.~V. Rakonjac, Y.-H. Chen, J.~J. Longdell, P.~Goldner, J.~P.~R.
  Wells, and M.~F. Reid.
\newblock Phys. Rev. Lett. \textbf{123}, 057401 (2019).

\bibitem{Mothkuri_zeeman_2021}
S.~Mothkuri, M.~F. Reid, J.-P.~R. Wells, E.~Lafitte-Houssat, P.~Goldner, and
  A.~Ferrier.
\newblock Phys. Rev. B \textbf{103}, 104109 (2021).

\bibitem{macfarlane1987}
R.~M. Macfarlane and R.~M. Shelby.
\newblock In A.~A. Kaplyanskii and R.~M. Macfarlane, editors, Spectroscopy of
  Solids Containing Rare Earth Ions,  (Elsevier Science Publishers B.V.,
  Amsterdam~1987).

\bibitem{dieke1964}
G.~H. Dieke and B.~Pandey.
\newblock J. Chem. Phys. \textbf{41}, 1952 (1964).

\bibitem{crosswhite1967}
H.~M. Crosswhite and H.~W. Moos.
\newblock In H.~M. Crosswhite and H.~W. Moos, editors, Optical Properties of
  Ions in Crystals,  (Interscience, New York~1967).

\bibitem{Carnall1989}
W.~T. Carnall, G.~L. Goodman, K.~Rajnak, and R.~S. Rana.
\newblock J Chem. Phys. \textbf{90}, 3443 (1989).

\bibitem{GoBi96}
C.~G{\"o}rller-Walrand and K.~Binnemans.
\newblock In J.~K.~A.~Gschneidner and L.~Eyring, editors, Handbook on the
  Physics and Chemistry of Rare Earths, volume~23, 121,  (North-Holland,
  Amsterdam~1996).

\bibitem{liu_electronic_2006}
G.~Liu.
\newblock In G.~Liu and B.~Jacquier, editors, Spectroscopic {{Properties}} of
  {{Rare Earths}} in {{Optical Materials}},  ({Springer Science \& Business
  Media~}2006).

\bibitem{Reid2016}
M.~F. Reid.
\newblock In J.~C.~G. Bunzli and P.~V. K., editors, Handbook on the Physics and
  Chemistry of Rare Earths, volume~50, 47--64,  (North Holland,
  Amsterdam~2016).

\bibitem{reid_transition_2006}
M.~F. Reid.
\newblock In G.~Liu and B.~Jacquier, editors, Spectroscopic {Properties} of
  {Rare} {Earths} in {Optical} {Materials},  (Springer Science \& Business
  Media~2006).

\bibitem{mujaji1992}
M.~Mujaji, G.~D. Jones, and R.~W.~G. Syme.
\newblock Phys. Rev. B. \textbf{46}, 14398 (1992).

\bibitem{martin1993}
J.~P.~D. Martin, T.~Boonyarith, N.~B. Manson, M.~Mujaji, and G.~D. Jones.
\newblock J. Phys.: Condens. Matter \textbf{5}, 1333 (1993).

\bibitem{kornienko1972}
L.~S. Kornienko and A.~A. Rybaltovskii.
\newblock Sov. Phys. Solid State \textbf{13}, 1785 (1972).

\bibitem{Boldyrev2019}
K.~N. Boldyrev, M.~N. Popova, B.~Z. Malkin, and N.~M. Abishev.
\newblock Phys. Rev. B \textbf{99}, 041105(R) (2019).

\bibitem{Shiddiq2016}
M.~Shiddiq, D.~Komijani, Y.~Duan, A.~Gaita-Ari{\~{n}}o, E.~Coronado, and
  S.~Hill.
\newblock Nature \textbf{531}, 348 (2016).

\end{thebibliography}

\end{document}